\documentclass[structabstract]{aa}
\usepackage{graphicx}
\usepackage{txfonts}
\begin{document}
\title{A polar+equatorial wind model for broad absorption line quasars:\\ I. Fitting the \ion{C}{iv} BAL profiles}

%   \subtitle{I. Fitting the BAL profiles}

   \author{B. Borguet\inst{1}\fnmsep\thanks{PhD. grant student of the Belgian National Fund for Scientific Research (F.N.R.S.)}
          \and D. Hutsem\'ekers\inst{1}\fnmsep\thanks{Senior research associate F.N.R.S.}
          }

\institute{Institut d'Astrophysique et de G\'eophysique, University of Li\`ege,
              All\'ee du 6 Ao\^ut 17, B-4000 Li\`ege\\
              \email{b.borguet@ulg.ac.be}}

   \titlerunning{Fitting the \ion{C}{iv} BAL profiles in quasars with a polar+equatorial wind model}
   \authorrunning{B. Borguet \& D. Hutsem\'ekers}

   \date{Received September 15, 1996; accepted March 16, 1997}

% \abstract{}{}{}{}{}
% 5 {} token are mandatory

  \abstract
   {Despite all the studies, the geometry of the wind at the origin of the blueshifted broad absorption lines (BAL)
 observed in nearly 20\% of quasars still remains a matter of debate.}
   {We want to see if a two-component polar+equatorial wind geometry can reproduce the typical BAL profiles observed in these objects.}
   {We built a Monte Carlo radiative transfer code (called MCRT) to simulate the line profiles formed in a polar+equatorial wind in which the photons, emitted from a spherically symmetric core are resonantly scattered. Our goal is to reproduce typical \ion{C}{iv} line profiles observed in BAL quasars and to identify the parameters governing the line profiles.}
   {The two-component wind model appears to be efficient in reproducing  the BAL profiles from the P Cygni-type profiles to the more complex ones. Some profiles can also be reproduced with a pole-on view. Our simulations provide evidence of a high-velocity rotation of the wind around the polar axis in BAL quasars with non P Cygni-type line profiles.}
   {}


   \maketitle

%
%________________________________________________________________

\section{Introduction}
 \label{lintro}

  Depending on the selection technique and the definition used, about 20\% to 30\% of the quasars detected in recent surveys show the presence of the broad absorption line (BAL) troughs associated with the emission lines in their rest frame UV spectrum (e.g. Knigge et al. \cite{kn08}, Ganguly et al. \cite{ga08}). These BALs, reminiscent of the P Cygni-type profiles seen in the spectra of massive stars, are mainly observed in high ionization lines like \ion{C}{iv} and \ion{Si}{iv} and are sometimes detected in lower ionization species like \ion{Mg}{ii}. They reveal strong outflows from quasars (Scargle \cite{sc72}), which can reach velocities up to 0.2 c (Foltz et al. \cite{fo83}).

   Despite the large number of observations, the physical and geometrical properties of the wind at the origin of the BALs remain largely unknown (e.g. Brotherton \cite{br07}). Moreover, the distance at which those objects are found ($z \geq 1.5$, so that \ion{C}{iv} is shifted in the optical domain) hampers direct observation of the regions at the origin of the BALs even with the best telescopes presently available. Thus, all the information we can get about the inner regions of BAL quasars comes mainly from indirect observations.

   The first attempts to model the BAL profiles considered the resonant scattering of photons emitted by a continuum source in a spherically symmetric stellar-like wind (e.g. Scargle et al. \cite{sc72}, Surdej \& Hutsem\'ekers \cite{su87}). However, the growing number of observed spectra displaying a huge variety of line profiles (Korista et al. \cite{ko93}) revealed the need for other wind models. Facing the diversity of line profiles, Turnshek (\cite{tu84a}) proposed that BAL quasars could be broadly divided into two samples: those quasars that exhibit smooth P Cygni-type profiles, and other ones that display an absorption trough that is detached in velocity from the associated weaker and wider emission peak. These observations indicate that the properties of the wind are more complex than the simple spherically  symmetric outflow inferred for stellar winds (Lee \& Blandford \cite{le97}). However, as emphasized by Turnshek (\cite{tu84b}), it is very likely that distinct types of BAL QSOs do not exist but are instead different manifestations of the same phenomenon.

   The similarities of the emission line, optical continuum, and infrared properties of BAL and non-BAL QSOs (e.g. Weymann et al. \cite{we91}, Gallagher et al. \cite{ga99}, Reichard et al. \cite{re03}, Gallagher et al. \cite{ga07}), as well as the spectropolarimetric observations (e.g. Schmidt \& Hines \cite{sc99}, Ogle et al. \cite{og99}, Lamy \& Hutsem\'ekers \cite{la04}), favor a unification by orientation scheme for the BAL QSOs over the evolutionary scheme (Hazard et al. \cite{ha84}, Becker et al. \cite{be00}). In the unification by orientation scheme, only a fraction (roughly corresponding to the observed fraction of BAL QSOs) of the continuum source is covered by optically thick material producing the broad absorption lines, which suggests a disk-like equatorial geometry for the BAL region (e.g. Turnshek \cite{tu84a}, Hamann et al. \cite{ha93}, Murray et al. \cite{mu95}, Elvis et al. \cite{el00}, Yamamoto \cite{ya02}). Such a geometry is supported by theoretical studies and commonly accepted, since the QSOs are thought to be powered by accretion of matter onto a supermassive black hole in the form of a disk, from which the wind could be launched. However, the recent discovery of radio loud BAL QSOs (e.g. Becker et al. \cite{be00}) and subsequent radio variability studies reveal polar outflows in at least some of them (Brotherton et al. \cite{br06}, Zhou et al. \cite{zh06}, Ghosh \& Punsly \cite{gh07}). Models combining polar and equatorial components have also been suggested (e.g. Lamy \& Hutsem\'ekers \cite{la04}) and evaluated from a theoretical point of view (Pereyra et al. \cite{pe04}, Proga et al. \cite{pr00}, Proga \cite{pr03}, Proga \& Kallman \cite{pr04})

   In this context and given the similarities between typical BAL profiles (e.g. Korista et al. \cite{ko93}) and the line profiles produced by a two component polar+equatorial wind like the one presented by Bjorkman et al. (\cite{bj94}), our goal in this first paper is to determine whether such a simple two-component wind can qualitatively reproduce the various types of line profiles observed among the BAL QSOs. We also try to identify the key ingredients needed to reproduce BAL profiles. In a second paper, we will investigate the effect of microlensing on these profiles, aiming at a realistic interpretation of the spectral differences observed in gravitationally lensed BAL QSOs like H1413+117 (cf. Hutsem\'ekers et al. \cite{hu09}).

In Sect.\ref{sec1}, we present MCRT, the Monte Carlo radiative transfer code we implemented in order to simulate resonance line profiles in a two-component axisymmetric wind. In Sect.\ref{parmstud} we briefly identify the influence of the wind model parameters on the line profiles we computed. In Sect.\ref{fitbal}, we show how MCRT is able to reproduce typical \ion{C}{iv} BAL QSOs line profiles. We discuss the results of the line profile fitting and summarize our conclusions in the last two sections of the paper.

%__________________________________________________________________

\section{The MCRT code}
\label{sec1}

MCRT is a Fortran77 fully 3D Monte Carlo (MC) radiative transfer (RT)
code that we built to compute the resonance line profiles produced in
axisymmetric winds. The use of the Monte Carlo simulation technique allows
the radiative transfer equation to be solved exactly (i.e. without
making use of the Sobolev approximation), as well as ensuring the self consistent treatment
of the radiative coupling between distant regions in a wind
subject to more complex velocity fields than monotonic radial laws
(e.g. Knigge et al. \cite{kn95}).

Monte Carlo RT code have been extensively described
(e.g. Knigge et al. \cite{kn95}, Wood et al. \cite{wo01}, Dijkstra et
al. \cite{di06}), so that we only recall here the fundamental principles of
this technique and the particularities of the code we developed.

As stated in the introduction, our main goal is to identify of the key ingredients (geometry and overall kinematics) of the wind governing the typical profile of the BAL QSO UV resonance lines. Thus we do not consider negligible effects, such as the relativistic
ones that remain small even for outflows with high ($\rm{v}_{max} \leq 0.2 c$) terminal speed (Hutsem\'ekers \& Surdej \cite{hu90}) or the fact that the line is a resonance doublet, because the velocity separation
of the doublet components is small with respect to $\rm{v}_{max}$ (e.g. Hewitt et al. \cite{he74}, Grinin \cite{gr84}).

\subsection{Radiative transfer with Monte Carlo techniques}

When using the MC technique, the solution of the RT equation is
found by following a huge number of photons on
their way through the wind. Each step in the photon's life (position and direction of emission,
position of interaction, etc) is
determined by the mean of random numbers distributed according to the normalized probability
density function (NPDF) of the corresponding simulated physical process.
Thus if one wishes that the frequency $\nu_i$ of all of the emitted photons
follows a given law $L(\nu)$ over the frequency interval $[\nu_{min},\nu_{max}]$,
then $\nu_i$ will be randomly chosen by solving the transformation equation
(Press et al. \cite{pr92}):
 \begin{equation}
   \label{transfo}
   \xi = \int_{\nu_{min}}^{\nu_{i}} L(\nu) d\nu \,,
 \end{equation}
where $\xi$ is a random number drawn from a uniform distribution in
the interval $[0,1]$. In MCRT this number is generated using the ``ran2" subroutine of
Press et al. (\cite{pr92}). In the following, each new
occurrence of $\xi$ refers to the call of such a new random number.
In general, there is seldom an analytical
solution to Eq.\ref{transfo} so we implemented the ``table lookup method"
(see Avery \& House \cite{av68}), which allows us arbitrary NPDF's.

In MCRT the initial position of emission of the photons is chosen isotropically
on the surface of the continuum emission region, which is modeled by a sphere of radius $R_{in}$
and of infinite optical depth ($\tau_C = \infty$) located at the center of the wind. The direction of
travel through the wind is then determined by randomly sampling a half sphere
taking into account that the photons are forced to leave the continuum source upward.

\subsection{Continuum photons and resonance scattering}

The wind is filled with 2-level atoms whose rest-frame normalized absorption profile $\phi_{abs}$ is
described by a Gaussian (Natta \& Beckwith \cite{na86}, Knigge et al. \cite{kn95})
such that
\begin{equation}
  \label{profab}
   \phi_{abs} (\nu-\nu_0) = \left\{ \begin{array}{cl} \phi_{abs} (\nu-\nu_0,\sigma_{turb}) & \mbox{if} ~|\nu-\nu_0| \leq |\Delta \nu_{abs}| \\ 0 & \mbox{everywhere else} \end{array} \right.  \,,
\end{equation}
where $\nu_0$ is the rest-frame frequency of the considered transition, and
\begin{equation}
  \label{gaussi}
  \phi_{abs} (\nu-\nu_0,\sigma_{turb}) = \frac{1}{H} \left \{ \frac{1}{\sqrt{\pi} \sigma_{turb}}~ \exp~\left(\frac{-(\nu-\nu_{0} )^{2}}{\sigma_{turb}^{2}} \right) +K \right \} \,,
\end{equation}
in which $K$ ensures a continuous transition between the absorption profile and
the zero intensity. Here, $H$ is a constant allowing for the normalization of the line profile
over the interval $\nu_0 \pm |\Delta \nu_{abs}|/2$, where $\Delta \nu_{abs}$ is the
full width at zero intensity (FWZI) of the absorption profile. We choose the value of $\Delta \nu_{abs}$
to ensure the continuity of the absorption profile at the border of the interval $\nu_0 \pm |\Delta \nu_{abs}|/2$.
The parameter $\sigma_{turb}=\Delta \nu_{turb}/(2 \sqrt{2 \ln{2}})$ is such
that $\Delta \nu_{turb}=2~\nu_0 (\rm{v}_{turb}/c)$ is the FWHM of the absorption
profile. The velocity $\rm{v}_{turb}$ includes the thermal and the macroscopic turbulence
components in the wind that broaden the absorption profile. We assume $\rm{v}_{turb}$ to be
constant throughout the wind.

Owing to the velocity field $\overrightarrow{\rm{v}}(r,\theta,\phi)$ present in the wind,
the initial frequency $\nu_i$ of a photon flying in the direction $\overrightarrow{n}$ is seen Doppler-shifted by an atom of the wind in such a way that its ``local" frequency $\nu_l$ in the atom rest-frame is given by
\begin{equation}
  \nu_{l} = \nu_i~\left(1- \frac{\overrightarrow{\rm{v}} ~ \overrightarrow{n}}{c} \right) \,.
\end{equation}
Thus a photon will enter in resonance with the surrounding atoms only if its local
frequency fulfills the condition defining the so-called ``resonance zone":
\begin{equation}
  \nu_0- \frac{\Delta \nu_{abs}}{2} < \nu_l < \nu_0+ \frac{\Delta \nu_{abs}}{2} \,.
\end{equation}
When the photon enters such a region, the opacity of the medium becomes
nonzero as does the probability of being absorbed. If $n$ resonance
zones are found along the direction of propagation of the photon in the
wind, the total optical depth $\tau_{tot}$ seen by the
photon until it escapes the wind is simply computed as
\begin{equation}
  \tau_{tot}(\nu_i) = \sum_{j=1}^{n} \int_{a_j}^{b_j} \kappa_{\nu_0} \phi_{abs} (\nu_l-\nu_0,\sigma_{turb})~ds \,,
\end{equation}
where $\kappa_{\nu_0}$ is the total absorption coefficient of the considered resonance transition, and $a_j$ and $b_j$
are respectively the coordinates of the beginning and of the end of the $j^{th}$ resonance zone found along the line of flight
of the photon.

Given the probabilistic interpretation of the RT, a photon experiencing a total optical depth
of $\tau_{tot}(\nu_i)$ has a probability $p=e^{-\tau_{tot}(\nu_i)}$ of escaping the medium without being absorbed.
This interpretation is used in the MC code to identify the occurrence and the position of
the scattering sites along the path of the photon through the wind. Indeed by using the transformation
equation (Eq.\ref{transfo}) we can determine the random optical depth $\tau_{MC}$ at which the
photon interacts :
\begin{equation}
 \tau_{MC} = - \ln (1-\xi)\,.
\end{equation}
 Because we stored the run of $\tau_{tot} = \tau_{tot}(s)$ along the photon path, it is easy to invert this relation and find the point $s$ where $\tau_{tot}=\tau_{MC}$. At this location the photon is radiatively absorbed and then instantaneously re-emitted at a frequency and in a direction chosen by assuming a complete redistribution in frequency and direction (CRFD, Lucy~\cite{lu71}, Mihalas et al.\cite{mi76}).
This photon may then either be re-absorbed somewhere else in the wind or leave it and be detected by one of the detector (spectrographs, imagers) located around the wind.

To decrease the simulation time in the case of non-spherically symmetric winds we make an intense use of the
advanced concepts of ``first forced interaction" (e.g. Cashwell \& Everett \cite{ca59}, Witt \cite{wi77}) and ``peeling off" (e.g. Yusef-Zadeh et al. \cite{yu84}, Wood \& Reynolds \cite{wo99}) where we follow a photon packet rather than a single photon.

We checked the validity of our MCRT code by comparing the line profiles
we obtained to line profiles computed with two traditional methods for spherical winds that
allow an exact integration of the transfer equation. These benchmarks
are the well-known SEI method of Lamers et al. (\cite{la87}) and the comoving
frame method of Hamann et al. (\cite{ha81}). We noted good
agreement between the general shape of the computed profiles,
 regardless of the considered turbulence ($\Delta F/F \leq 5 \%$ on the normalized emission peak flux, as well as a good match between the absorption profiles). We also tested MCRT in the case of
axi-symmetric winds by comparing the profiles obtained with MCRT to those produced by
the SEI method adapted by Bjorkman et al. (\cite{bj94}). Once again
we observed good agreement between the line profiles produced by both methods (see Borguet \cite{bo09} for details).

\subsection{Pure emission}

While the shape of the \ion{C}{iv} line in BAL QSOs can be mostly governed by resonance scattering (Scargle et al. \cite{sc72}),
the presence of \ion{C}{iii}$]$ emission constitutes evidence that part of the emission is due to collisional
excitation (Turnshek \cite{tu84a}, Turnshek \cite{tu88}, Hamann et al. \cite{ha93}). To account for this second source of photons,
we allow the production directly in the wind of a fraction of photons $f_e=I_{pure~emission}/I_{continuum}$. The choice of the
location of the emission $(r_e, \theta_e, \phi_e)$ of these photons is made using a random sampling of the corresponding NPDF:
\begin{equation}
 \xi = p(r_e, \theta_e, \phi_e) = \int_{0}^{\phi_e} \int_{0}^{\theta_e} \int_{R_{in}}^{r_e} \eta(r,\theta,\phi) r^{2} \sin{\theta}~dr d\theta d\phi  \,,
\end{equation}
where the $\eta(r,\theta,\phi)$ is a function that describes the emissivity throughout the wind.
Once again, our goal here is not to provide a detailed self-consistent model of the wind so we
choose as a first guess an emissivity function of the form
\begin{equation}
  \eta(r,\theta,\phi) = n(r,\theta,\phi) \left( \frac{R_{in}}{r} \right)^{\gamma} \,,
\end{equation}
where $n(r, \theta, \phi)$ is the density of the ion through the wind and where the second term allows
taking the temperature distribution and the ionization fraction into account. In the
following we simply take $\gamma=1$ in order to reduce the number of free parameters.

Each photon emitted then makes its way through the wind where it can be scattered
and then finally escapes the wind in order to be detected by a distant observer.

\subsection{The wind model}

\subsubsection{\textbf{Introduction}}

As stated in the introduction, it is difficult to give a simple explanation
to the observed BAL profiles when using a spherically symmetric expanding wind. The obvious next type of geometry that can then be considered is the axi-symmetric one.
The simplest of these models would still consist in a wind originating from the central core.
Such a generic model, with polar and equatorial components like the
one presented in details by Bjorkman et al. (\cite{bj94}) produces line profiles
remarkably similar to those observed in some BAL QSOs.  Although
simple, this model is versatile enough to produce a variety of line
profiles, as observed in BAL QSOs. It constitutes a first good
approximation to the more complex wind from disk models proposed for
AGN outflows in which the BALR and the BELR are generally cospatial (Sect. \ref{lintro}).

In our model, we adopt stellar wind laws to describe the kinematics of
the winds observed in quasars. Indeed, quasar winds are also supposed
to be driven by radiation (Arav \& Li \cite{ar96}, Murray et al. \cite{mu95}) as
suggested by the line-locking in the spectra of some BAL
QSOs (e.g. Weymann et al. \cite{we91}, Korista et al. \cite{ko93}, Arav \cite{ar96,ar97}). However, there are important differences between stars and
quasars (e.g. Arav \cite{ar94}). One of them is the so-called overionization
problem caused by the strong UV/X-ray central source in quasars
(e.g. Proga et al. \cite{pr00}). Several scenarios have been suggested to
solve this problem: Murray et al. (\cite{mu95}), Murray \& Chiang (\cite{mu97}), Risaliti
\& Elvis (\cite{ri09}), Punsly (\cite{pu99}) and Ghosh \& Punsly (\cite{gh07}). In our study we assume the existence of shielding
material between the radiation source and the outflow that prevents the total ionization of the outflow (see Krolik \cite{kr99}).

Another difference between stellar objects and quasars comes from a significant
fraction of the radiation in quasars supposedly being emitted from an accretion disk rather than from a
spherically symmetric photosphere (e.g. Proga et al. \cite{pr00}). Outflows
with axial geometries have been studied by several authors, who show
that the flow can be launched vertically from the disk and then pushed
away by the radiation from the central source (Murray et
al. \cite{mu95}, Proga et al. \cite{pr00}) with the elevation of the wind
over the disk is still small when the flow starts to expand radially.
Since the launching conditions are unclear (Risaliti \& Elvis \cite{ri09})
and since we assume the BALR and the BELR to be cospatial, we simplify
the geometry by assuming that the wind is purely radial, which is
correct at some distance from the disk. We considered a wind launched
from a sphere with a radius equal to the inner radius of a typical
BELR (10$^{-2}$ pc). The outer radius of the BELR/BALR was chosen to be
1 pc, the radius at which the wind reaches the terminal velocity. If
needed, the anisotropy of the radiation field from the continuum
source can be readily introduced into our model.

%                                     One column figure (place early!)
%______________________________________________ Gamma_1 (lg rho, lg e)
   \begin{figure}
   \centering
   \includegraphics[width=8.5cm]{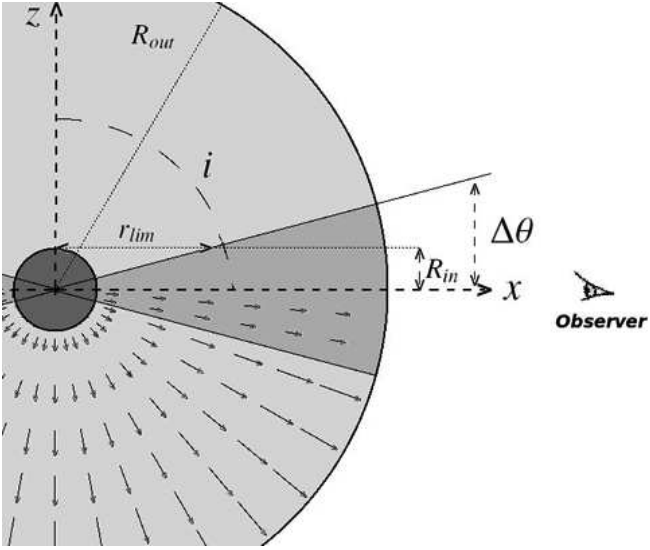}
      \caption{Illustration in the $x-z$ plane of the two-component wind used in our MC simulations, the velocity field is pictured by arrows in the lower part of the picture. The equatorial wind is shown edge-on (in dark grey) so that the wind model is rotationnaly symmetric along the $z$ axis. The viewing angle of the observer is $i=90 \degr$, and the absorption profile is produced by both component, the polar one contributing only at low velocities (i.e. for $x<r_{lim}$) given the disk opening angle and the viewing angle. Inspired from Fig.~8 of Bjorkman et al. (\cite{bj94}).}
         \label{windmod}
    \end{figure}

\subsubsection{\textbf{Velocity field and density law}}

Keeping in mind the aforementioned simplifications we then decided to implement the model
described by Bjorkman et al. (\cite{bj94}). We recall its basic
characteristics here.
The velocity field $\overrightarrow{\rm{v}}(r,\theta,\phi)$ considered here can be written in spherical coordinates:
\begin{equation}
  \overrightarrow{\rm{v}}(r,\theta,\phi) =  \rm{v}_r(r,\theta)~\overrightarrow{e_r} + \rm{v}_{\phi}(r,\theta)~\overrightarrow{e_{\phi}} + \rm{v}_{\theta}(r,\theta)~\overrightarrow{e_{\theta}} \,,
\end{equation}
where for simplicity we assume, as in Bjorkman et al. (\cite{bj94}), that the streamlines lie on
surfaces of constant polar angle so that $\rm{v}_{\theta} = 0$. The radial component of the velocity field
has the typical $\beta-$law shape:
\begin{equation}
 \label{veloradial}
  \rm{v}_r(r,\theta) = \rm{v}_{min}+(\rm{v}_{max}(\theta) - \rm{v}_{min}) \left ( 1-\frac{R_{in}}{r} \right )^{\beta} \,,
\end{equation}
where $\rm{v}_{min}$ is the wind speed at the surface of the source of the continuum. The variable terminal speed $\rm{v}_{max}(\theta)$ allows for the introduction of a slow expanding
equatorial wind in a faster polar wind:
\begin{equation}
  \rm{v}_{max}(\theta)= \rm{v}_{max}^{po}+(\rm{v}_{max}^{eq}-\rm{v}_{max}^{po}) f(\theta) \,,
\end{equation}
where
$\rm{v}_{max}^{eq}$ and $\rm{v}_{max}^{po}$ are respectively the terminal velocity of the equatorial and
the polar components. The function that allows the smooth transition between the two components of the wind
takes the form
\begin{equation}
\label{fteta}
 f(\theta) = 0.5 \left [ 1 + \frac{2}{\pi} \arctan{ \left ( \frac{sin{\Delta \theta} - \cos{\theta}}{\Delta \theta_t} \right ) } \right ] \,,
\end{equation}
where $\Delta \theta$ is the equatorial wind half opening angle and where the parameter $\Delta \theta_t$ controls the transition between the
equatorial and the polar component of the wind. We chose, as in to Bjorkman et al. (\cite{bj94}), a low value for $\Delta \theta_t = 0.001$ in order to keep this region much thinner than the equatorial wind opening angle.

The $\rm{v}_{\phi}$ component of the velocity field is simply given by assuming
conservation of the angular momentum in the wind:
\begin{equation}
 \rm{v}_{\phi} = \frac{\rm{V}_{rot}~R_{in}}{r} \sin{\theta} \,,
\end{equation}
where $\rm{V}_{rot}$ is the rotational speed at the surface of the source of the continuum.

The law governing the distribution of the ion density is derived from the
equation of continuity of matter in a central wind and is parameterized using
\begin{equation}
 n(r,\theta,\phi) = n_{0}(\theta) \left ( \frac{r}{R_{in}} \right )^{- \alpha} \left ( \frac{\rm{v}_r(r,\theta)}{\rm{v}_{min}} \right )^{-1} \,,
\end{equation}
with the parameter $\alpha$ implicitly including the radial variation of the ionization
(i.e. $\alpha=2$ meaning a constant ionization throughout the wind)
and $n_{0}(\theta)$ allowing the transition between the polar and the equatorial components:
\begin{equation}
  n_0(\theta) = n_0^{po} + (n_0^{eq} - n_0^{po}) f(\theta)\,,
\end{equation}
where $n_0^{eq}$ and $n_0^{po}$ are, respectively, the density of the considered ion at the base of the wind in the equatorial or in the polar component. The value of $n_{0}^{po}$ is
computed by specifying the value $\tau_{tot_p}$ of the total optical depth of the wind along the polar axis (i.e. $\theta=0\degr$) integrated over frequency:
\begin{equation}
  \tau_{tot_p} = \int_{-\Delta \nu_l}^{\Delta \nu_l} \tau_{tot}(\nu_i) d\nu_i \,,
\end{equation}
with $2~\Delta \nu_l$ the measured width of the observed line profile.
The value of $n_{0}^{eq}$ is fixed by the free parameter $k_{pm}$, which defines the ratio
of the ionic density between the two components at the base of the wind:
\begin{equation}
n_{0}^{eq}=k_{pm}~n_{0}^{po}\,.
\end{equation}

\section{Parameter study}
\label{parmstud}

Here we concentrate on the main parameters affecting the line profiles
 for a two-component wind where we suppose no pure emission (i.e. $f_e=0$). We do not discuss the  effects of the parameters governing
the velocity and density laws or the effects of the turbulent component in the wind since they are similar to those observed in the
well-understood spherically expanding wind case (Castor \& Lamers \cite{ca79}, Beckwith \& Natta \cite{be87}, Hamman \cite{ha81}, Lamers
et al. \cite{la87}).

The most important parameters we have to specify for
computing a line profile are the frequency integrated polar optical depth $\tau_{tot_p}$,
the ratio between the equatorial and the polar ionic density $k_{pm}$ , the velocity
ratio between the polar and the equatorial terminal speed $\rm{v}_{max}^{eq}/\rm{v}_{max}^{po}$, the disk half-opening angle $\Delta \theta$,
the viewing angle $i$, and the ratio $\rm{V}_{rot}/\rm{v}_{max}^{po}$ between the rotational speed of the source of continuum and the polar terminal velocity. The parameters used for the reference line profile are summarized in Table~\ref{table1}.
Such a parameter study was previously carried out by Bjorkman et al. (\cite{bj94})
using an SEI-type method. Because their calculations agree with ours using an
exact method, we only recall here the general effects produced by each parameter and refer the reader
to the Bjorkman et al. paper for further details. We do, however,
emphasize the effect of the wind rotation given
the important changes in the line profile produced by the
variation of this parameter, more particularly when the equatorial
disk is viewed near edge-on (Mazzali \cite{ma90},
Petrenz \& Puls \cite{pe96}, Busche \& Hillier \cite{bu05}).

%                                     Two column figure (place early!)
%______________________________________________ Gamma_1 (lg rho, lg e)
   \begin{figure*}
   \centering
   \includegraphics[width=15.4cm]{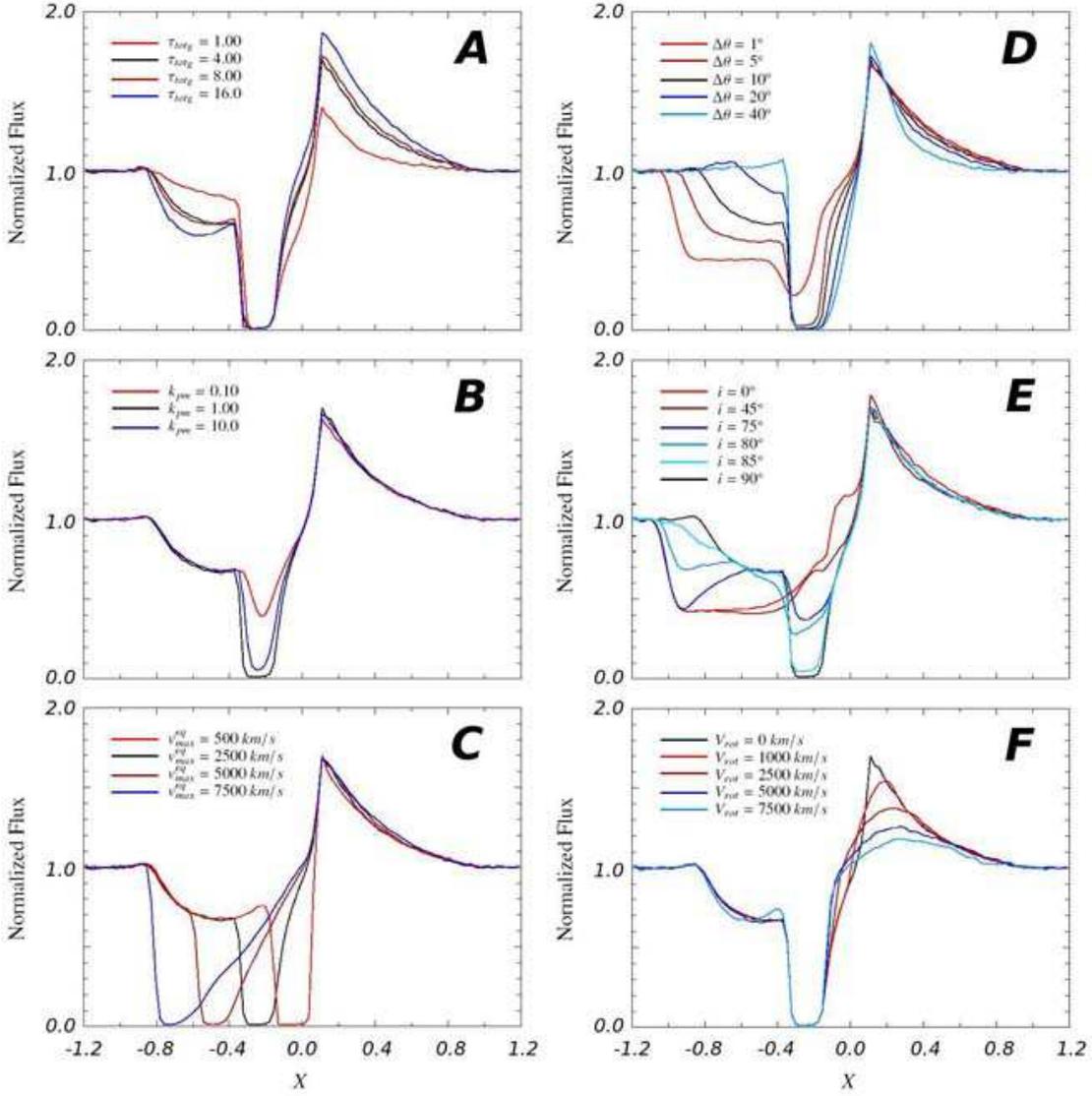}
      \caption{Illustration of the effects of the wind's main parameter on the
               line profile (see text for details). Each spectrum
               is represented as a function of the normalized wavelength $X=(\lambda-\lambda_0)/(\lambda_{max}-\lambda_0)$ and is the result of the simulation of $2~10^6$ photon paths through the wind ($S/N > 100$ on the continuum level).}
         \label{param}
    \end{figure*}
%

%_____________________________________________________________
%                                             Simple A&A Table
%_____________________________________________________________
%
\begin{table}
\caption{Wind parameter for the benchmark model}   % title of Table
\label{table1}      % is used to refer this table in the text
\centering                          % used for centering table
\begin{tabular}{l l | l l}        % centered columns (4 columns)
\hline\hline                 % inserts double horizontal lines
Parameter & Value & Parameter & Value\\    % table heading
\hline                        % inserts single horizontal line
   $\rm{v}_{max}^{po}$ &  $10000$ $\rm{km~s}^{-1}$ & $\alpha$&         $2.00$\\      % inserting body of the table
   $\rm{v}_{min}$ &       $100$ $\rm{km~s}^{-1}$  &   $\beta$&          $1.00$\\
   $\rm{v}_{max}^{eq}$&   $2500$ $\rm{km~s}^{-1}$  &  $\Delta \theta$&  $15\degr$\\
   $\rm{V}_{rot}$&        $0$    $\rm{km~s}^{-1}$ &    $\Delta \theta_t$&$0.001$ \\
   $\rm{v}_{turb}$&       $500$  $\rm{km~s}^{-1}$ &    $i$&              $90\degr$ \\
   $\tau_{tot_p}$ &  $4.00$ &    $f_e$& $0.0$\\
   $k_{pm}$&         $10.0$ & $\gamma$& $1.0$\\

\hline                                   %inserts single line
\end{tabular}
\end{table}

The profiles illustrated in the panel A of Fig.\ref{param} are representative of the profiles
that can be produced in a two-component wind when the equatorial component is seen near edge-on.
The shape of these profiles is constituted by typical P Cygni-type profile extending to the higher
velocities coming from the polar component, but where a sharp absorption trough is produced at lower velocities by the slowly expanding equatorial wind.
In this panel we illustrate the evolution
of the line profiles as a function of the
total polar optical depth integrated over frequencies $\tau_{tot_p}$. Similar to the spherically symmetric expanding wind case (e.g. the atlas constructed by
Castor \& Lamers \cite{ca79}), when $\tau_{tot_p}$ is increased, we observe an increase in the emission peak and in the
equivalent width of both the emission and absorption parts of the line profile. No variations are observed in the equatorial absorption
since that part of the profile is already saturated for the lower value of $\tau_{tot_p}$.

The effect of the $k_{pm}$ ratio is illustrated in the panel B of Fig.~\ref{param}. When $k_{pm}$ is decreased,
the depth of the absorption profile of the equatorial component evolves in the same way. As in Bjorkman et al. (\cite{bj94}),
we observe that the equatorial component is still optically thicker to the radiation than the polar component
even for $k_{pm} = 0.1$, and this because of the narrower velocity range of the equatorial component.

The change in the ratio between the teminal speed $\rm{v}_{max}^{eq}$ and $\rm{v}_{max}^{po}$ of the two components
essentially modifies the position of the edge of the low-velocity absorption component (see panel C of Fig.~\ref{param}). We also observe
that, when $\rm{v}_{max}^{eq}$ is increased, the absorption due to the presence of the disk no
longer remains black near the line center. Indeed, to produce a black absorption, the disk has to cover
the whole source of continuum. But for an equatorial wind with a half-opening angle $\Delta \theta$, this
only happens beyond the radial coordinate $r>r_{lim}=R_{in}~\rm{cotan}{\Delta \theta}$ (see Fig.~\ref{windmod}), which corresponds
to the velocity $\rm{v}_{lim}^{eq} \sim \rm{v}_r(r_{lim},90\degr$), the value of which is sensitive to $\rm{v}_{max}(\theta)$ (cf. Eq. \ref{veloradial}).

In panel D of Fig.~\ref{param}, we illustrate the evolution
of the line shape as a function of the half-opening angle
of the equatorial wind $\Delta \theta$. We observe the development of a broad
high-velocity absorption when $\Delta \theta$ is decreased. This observation is
once more explained by considering the radial coordinate $r_{lim}$ at which the disk
completely covers the source of the continuum. As the high-velocity absorption (due to the polar component)
occurs before the equatorial wind, the high absorption velocity range is limited within the interval $[0,\rm{v}_{lim}^{po}]$,
where $\rm{v}_{lim}^{po} \sim \rm{v}_r(r_{lim},0\degr$). In contrast, when $\Delta \theta$ is increased, we
observe the apparition of a secondary blueshifted emission peak beyond $\rm{v}_{max}^{eq}$ given
that, for a sufficiently wide disk, there is no remaining polar absorption of the
continuum.

The viewing angle $i$ to the wind plays a critical role in the line profile (see panel E of Fig.~\ref{param}).
Indeed the equatorial absorption component is only seen when the disk is viewed near edge-on
(i.e. $i=90\degr$). Because $i$ is decreased from an edge-on to a pole-on ($i=0\degr$) view of the disk,
the low-velocity absorption produced by the disk decreases, since only the high-velocity end of
this component covers the continuum.

   \begin{figure}
   \centering
   \includegraphics[width=8.5cm]{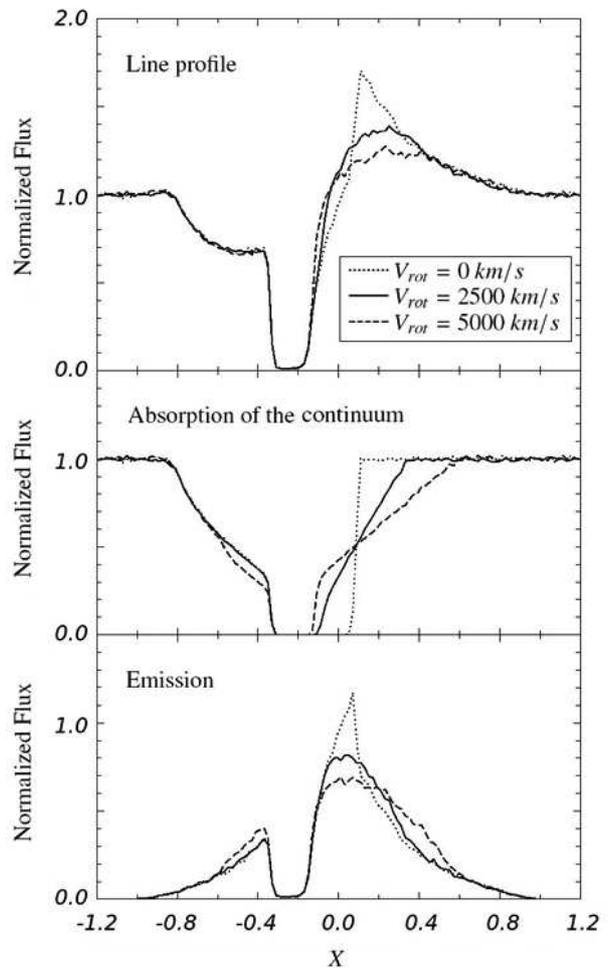}
      \caption{Closer illustration of the effect of the rotation of the wind over the line profile,
        its part in emission, and its part in absorption. The dotted line represents the profiles for which
        the rotational velocity is zero, while the straight line represents the profiles with $\rm{V}_{rot}=0.25~v_{max}^{eq}$,
        and the dashed line a wind where $\rm{V}_{rot}=0.50~v_{max}^{eq}$. The parameter used are those gathered in Table~\ref{table1} (see text).}
         \label{figredabs}
    \end{figure}

Finally, panel F of Fig.~\ref{param} illustrates the modifications of the line profile
when varying the rotational speed of the wind. The major effect of the rotation is observed
on the emission peak, which decreases and is significantly redshifted when the $\rm{V}_{rot}/\rm{v}_{max}^{po}$ ratio
is increased. Indeed, when there is no rotation, the emission peak is relatively sharp because
that part of the profile is produced in the inner, optically thicker regions of the wind, whose
velocity range is narrow. When rotation is considered, the resonance zones become twisted
in such a way that the inner regions are distributed over a wider range of projected velocity. This induces
the smoothening of the emission peak since the rotation allows the absorption to occur in the redder
part of the continuum at frequencies centered on the rest-frame frequency of
the line transition $\nu_0$ (e.g. $X=0.0$) (see Fig.\ref{figredabs} and also Hall et al. \cite{ha02} for a detailed explanation). The bluer part of the absorption profile is not significantly affected by the
rotation because this part of the profile is produced in front of the continuum source, i.e.
where the projected rotational velocity is nearly zero.

\section{Fitting BAL profiles with the \textbf{polar+equatorial wind} model}
\label{fitbal}

   \begin{figure*}
   \centering
   \includegraphics[width=15.4cm]{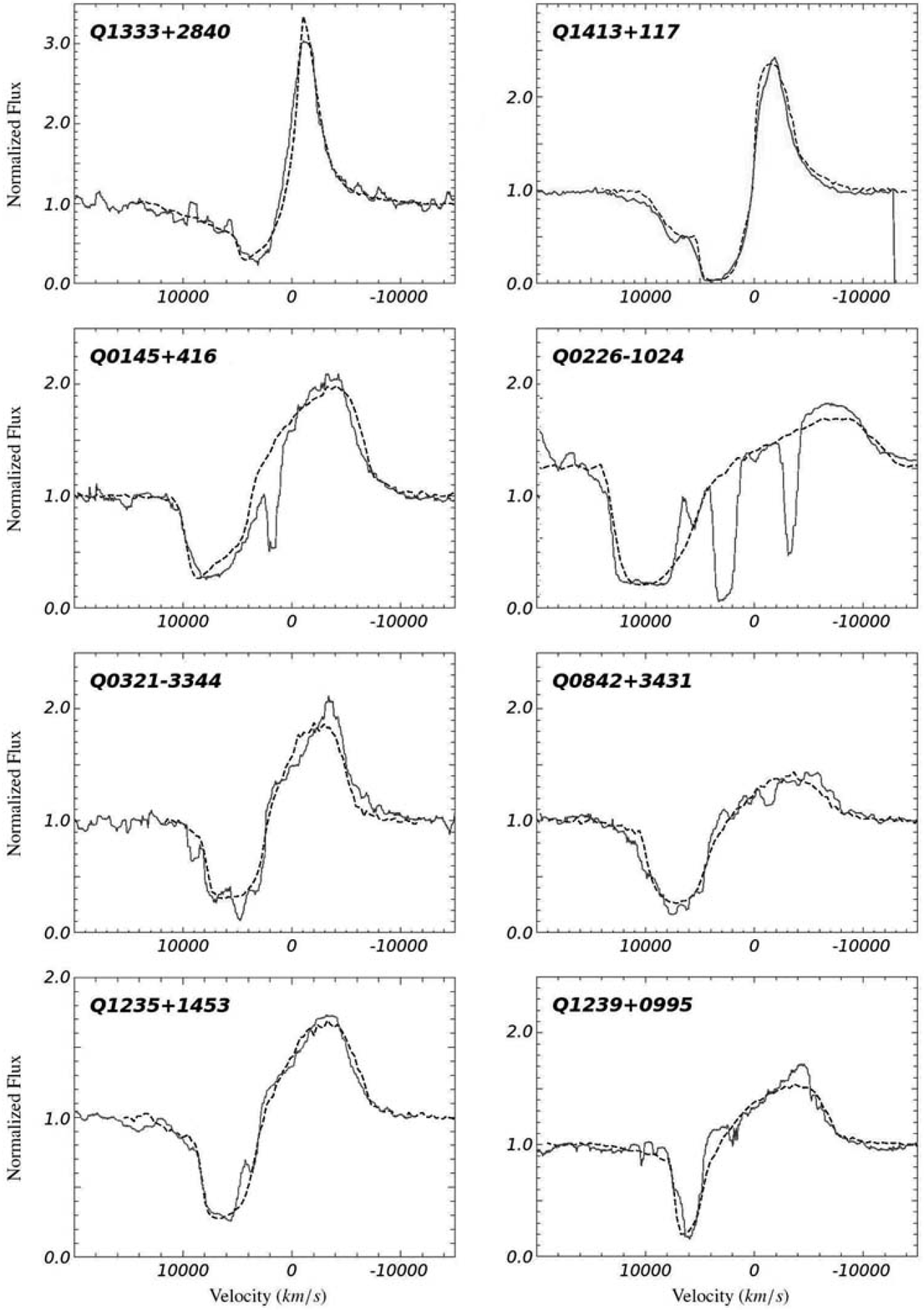}
      \caption{Illustration of the best fit for eight out of the ten \ion{C}{iv} line profile selected from
the BAL quasar sample of Korista et al. (\cite{ko93}). The observed spectrum is represented by a full line and
the two-component model profile by a dashed line. Each spectrum is normalized to the continuum level
and represented as a function of the velocity.}
         \label{fite1}
    \end{figure*}

   \begin{figure*}
   \centering
   \includegraphics[width=15.4cm]{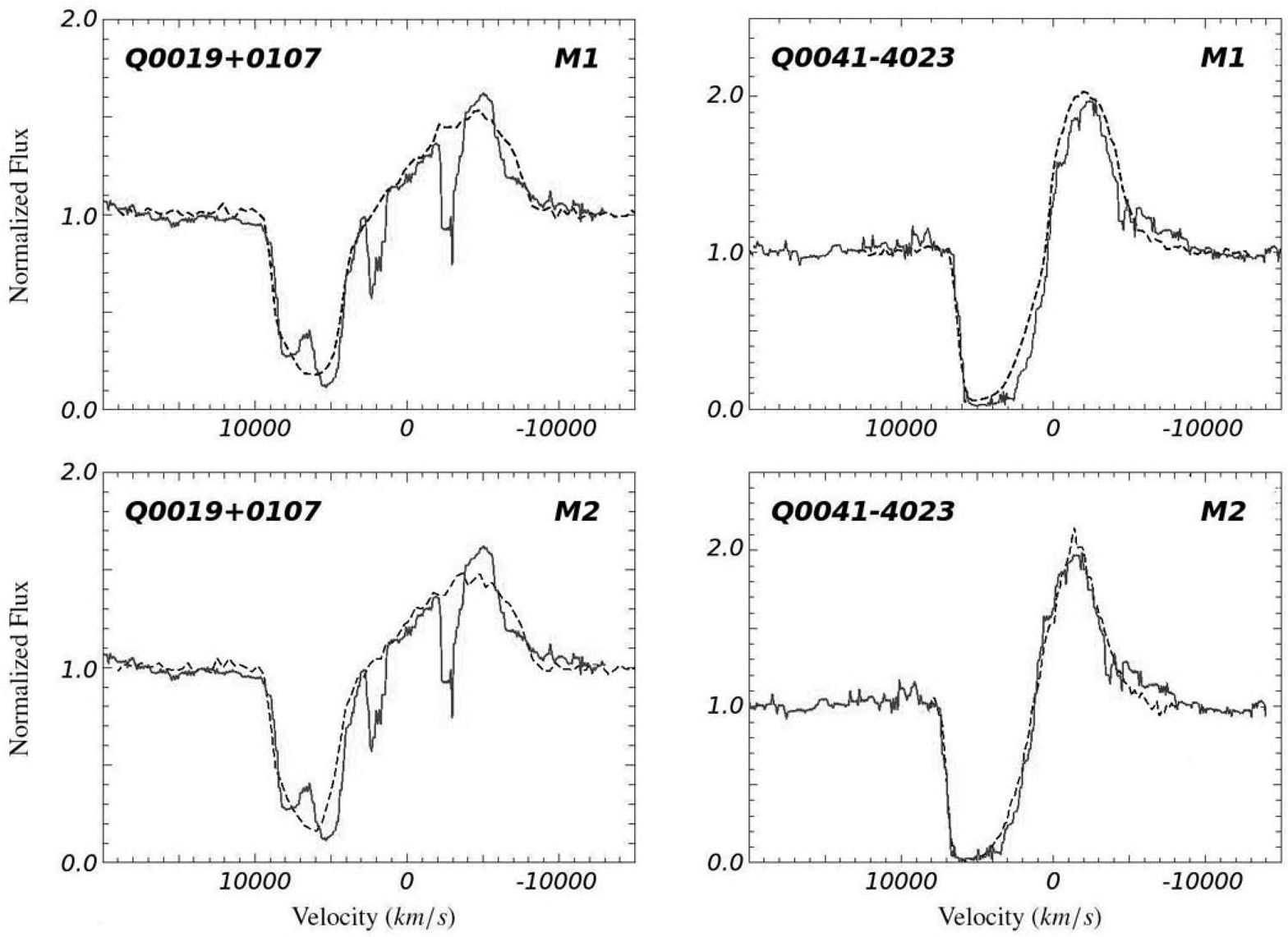}
      \caption{Illustration of the non-uniqueness of the fit for two particular objects of our sample. The upper panels (labelled M1) of this figure show that the \ion{C}{iv} line profile of these two objects can be fitted in the framework of the two-component wind model. In the lower panels (labeled M2), we show that the lack of typical two-component wind signatures in the line profile allows alternative fits with different parameters (see text).
      As in Fig. \ref{fite1} the observed spectrum is represented by a full line and the two-component model profile by a dashed. Each spectrum is normalized to the continuum level and represented on a velocity scale.}
         \label{fite2}
    \end{figure*}

In this section we try to reproduce a representative sample of \ion{C}{iv} line profiles observed in BAL QSOs and extracted from the paper of Korista et al. (\cite{ko93}), which provides homogeneous and high-quality spectroscopic data for a large sample of BAL QSOs.
Our goal is to check that such profiles can be modeled using a simple two-component
wind geometry for the BALR. Ten spectra were carefully selected out of
the 72 BAL QSOs presented in the Korista et al. paper on the
basis of two criteria.
\begin{itemize}
\item{}We retained as far as possible objects free of doublet effects (i.e. $\rm{v}_{max}^{po} \gg$ \ion{C}{iv} doublet separation), non-blended and with smooth emission/absorption \ion{C}{iv} line profiles (i.e those less affected by multiple absorption troughs).
\item{}We selected objects for which the continuum can be securely defined on both sides
of the \ion{C}{iv} line profile, allowing for a correct normalization of the spectra.
\end{itemize}

The observed and the fitted spectra of each selected object are illustrated,
normalized to the continuum level, in Figs.~\ref{fite1} and \ref{fite2}. We note good qualitative agreement between
the observed line profile and the model. The parameters
used in the fit of each \ion{C}{iv} line profile are collected in Table~\ref{tabfit}.
While some parameters ($\rm{v}_{max}^{eq}$, $\rm{v}_{max}^{po}$, $i$, $\tau_{tot_p}$, $\alpha$, $f_e$...) were allowed to vary, we tried to keep fixed the radial velocity gradient in
the wind (the $\beta$ parameter), the emission parameter $\gamma$ and the disk opening angle
$\Delta \theta$. The latter was chosen equal to the value generally adopted for the opening angles of disk winds in AGN models (Murray et al. \cite{mu95}, Proga \& Kallman \cite{pr04}). However, even those parameters were varied in a few cases to slightly modify the part of the profile in emission.
In particular, we found that the opening angle $\Delta \theta$ of the disk has to be decreased for fitting the high-velocity absorption component (see Sect.~\ref{parmstud})  observed in the P Cygni-type line profiles (Q1333+2840 and Q1413+1143).

\begin{table*}
\begin{minipage}[t]{\textwidth}
\caption{Parameter of the two-component wind model for the best fit of the \ion{C}{iv} BALs in the selected
quasar sample.}
\label{tabfit}
\centering
\renewcommand{\footnoterule}{}
\begin{tabular}{l r l r r r r l l l l l l}     % 12 columns
\hline\hline
BAL QSO Name & $ z $ & $\tau_{tot_p}$ & $k_{pm}$ & $\rm{v}_{max}^{po}$ ($\rm{km~s}^{-1}$) & $\rm{v}_{max}^{eq}$ ($\rm{km~s}^{-1}$) & $\rm{V}_{rot}$ ($\rm{km~s}^{-1}$) & $\alpha$ & $\beta$ & $\Delta \theta$ ($\degr$) & $i$ ($\degr$) & $f_e$ & $\gamma$\\
\hline
 Q1333+2840 & 1.897 & 3.00 & 10.0 & 13000 & 4000 & 1000 & 3.0 & 0.8 & 10 & 80 & 0.21 & 1.0 \\
 Q1413+1143 & 2.522 & 10.0 & 15.0 & 11000 & 3500 & 2000 & 3.0 & 0.8 & 10 & 88 & 0.16 & 1.0 \\
 Q0145+0416 & 1.990 & 1.00 & 12.0 & 18000 & 9600 & 6400 & 2.5 & 1.2 & 15 & 80 & 0.18 & 1.0 \\
 Q0226-1024 & 2.164 & 1.00 & 10.0 & 25000 & 12500 & 12500 & 2.0 & 1.5 & 15 & 87 & 0.10 & 1.0 \\
 Q0321-3344 & 1.955 & 10.0 & 10.0 & 10000 & 7500 & 5200 & 2.5 & 1.5 & 15 & 75 & 0.16 & 1.0 \\
 Q0842+3431 & 2.100 & 14.0 & 5.00 & 16000 & 10000 & 8000 & 3.0 & 1.5 & 15 & 78 & 0.04 & 1.0 \\
 Q1235+1453 & 2.835 & 5.00 & 10.0 & 13000 & 8000 & 7000 & 2.5 & 1.5 & 15 & 76 & 0.12 & 1.0 \\
 Q1239+0995 & 1.970 & 3.00 & 5.00 & 16000 & 8000 & 8000 & 3.0 & 1.0 & 15 & 82 & 0.10 & 1.0 \\
 Q0019+0107 M1 & 2.075 & 1.00 & 15.0 & 16000 & 9000 & 8000 & 2.5 & 1.5 & 15 & 90 & 0.11 & 1.0 \\
 Q0019+0107 M2 & 2.075 & 0.00 & 15.0\footnote{$k_{pm}$ represents the
total equatorial ($i=90 \degr$) optical depth integrated over frequencies since $\tau_{tot_p}=0$.} & - & 9000 & 8000 & 2.5 & 1.5 & 15 & 90 & 0.11 & 1.0 \\
 Q0041-4023 M1 & 2.450 & 5.00 & 10.0 & 10000 & 5000 & 3000 & 2.0 & 1.5 & 15 & 90 & 0.18 & 1.0 \\
 Q0041-4023 M2 & 2.450 & 15.0 & 5.0 & 6000 & 3000 & 3000 & 2.0 & 0.5 & 15 & 0 & 0.07 & 1.0 \\

\hline
\end{tabular}
\end{minipage}
\end{table*}

The sample contains various types of BAL spectra, from the
P Cygni-type to the more complex profiles discussed in the paper of
Korista et al. (\cite{ko93}). This subdivision of BAL QSO spectra was
introduced by Turnshek (\cite{tu84b}) and based on the visual properties of the BAL
profiles, where the P Cygni-type profiles with a smooth absorption and a high emission peak relative to the continuum level are distinguished from the complex/detached absorption ones with a lower emission peak.
However we tried to model all these line profiles using the same wind geometry
since, as suggested by Turnshek (\cite{tu84b}), it is likely that distinct types of BAL profiles
are the consequence of a unique mass-loss phenomenon.

The empirical fitting procedure we adopted here makes the distinction between three types of line profiles, each possessing its own set of spectral signatures. The first kind of line profile is the P Cygni-type (Q1333+2840 and Q1413+117 see top panels of Fig.\ref{fite1}). In this subsample, the absorption trough seems constituted of two subtroughs covering two overlapping velocity ranges, i.e. an optically thick ``narrow" ($\sim 4000$ $\rm{km}~s^{-1}$ wide) absorption component superimposed on an optically thin absorption trough extending up to velocities of 10000 $\rm{km~s}^{-1}$. These characteristics suggest that there is an equatorial denser disk-like region seen nearly edge-on and producing the narrow absorption observed. The necessity of considering such a more slowly expanding disk in a polar wind is illustrated particularly well when observing the \ion{C}{iv} line profile of Q1413+117. In this object, the absorption profile extends to velocities up to -10000 $\rm{km}~s^{-1}$, which implies, through resonance scattering, the
presence of an underlying emission component extending from -10000 $\rm{km}~s^{-1}$ to
10000 $\rm{km}~s^{-1}$. Since the flux is almost zero at slower
velocities ($\rm{v} \ge -4000$ $\rm{km}~s^{-1}$), a part of these photons has to
be re-absorbed somewhere, suggesting there is an optically thicker region
on the line of sight to the observer. This observation fits the framework
in which a part of the emission from the BELR is absorbed in the BALR (e.g. Turnshek et al. \cite{tuetal88},
Hamann et al. \cite{ha93}).

The second type of line profile we considered is presented in the lower panels of Fig. \ref{fite1}.
These profiles are characterized by an asymmetric emission peak whose intensity relative to the continuum level is
significantly lower than in the P Cygni-type ones. Some individuals of this subsample also show, similar
to the P Cygni-type ones, higher velocity absorption through superimposed to a
narrower component (best seen in, e.g., Q0842+3431, Q1235+1453). The introduction of a significant rotation of the wind around
the polar axis is required to produce the broad asymmetric emission peaks observed in the \ion{C}{iv} line profile of seven out of the ten
objects in our sample (Q0019+0107, Q0145+416, Q0226-1024, Q032-3344, Q0842+3431, Q1235+1453, and Q1239+0995). The rotational speed generally
needs to be quite higher than the polar terminal velocity (reaching up to $50$ \% of $\rm{v}_{max}^{po}$) and is approximately equal to
the equatorial wind terminal velocity. For the P Cygni-type profiles (Q1333+2840, Q1413+1143 or even Q0041-4023),
it is also necessary to consider a rotation of the wind around the polar axis, although it must be smaller to
properly account for the higher emission peak and the sharp transition between the emission and the absorption parts of the profiles.

In each case, the best fit was chosen by eye, once the model was qualitatively similar
to the profile of the observed line. The simulation time needed to produce a single line
profile in these optically thick winds prevents us from using a $\chi^2$ type technique while searching for the best model. However, this is not a main drawback since our main goal is to show that
a simple wind model is able to approximately reproduce a variety of resonance line profiles observed in
BAL QSOs. Moreover, given the degeneracy between some of the model parameters ($k_{pm}$ with $i$,
$\rm{v}_{max}^{eq}$ with $i$, $\beta$ with $\alpha$, etc), as well as the difficulty in some cases of correctly evaluating
$\rm{v}_{max}^{po}$ or other model parameters, more than one best fit is generally possible.

The absence of the two-component wind signatures defined above in the \ion{C}{iv} line profile of some objects led us
to define a third subsample of line profiles. These line profiles can be fitted
by a two-component wind (see upper panel of Fig.~\ref{fite2}). However, some profiles (the
prototype being Q0019+0107) do not show evidence of a polar absorption component at high velocities,
suggesting that the polar outflow may not be present in these objects. Several tests performed on such
line profiles with MCRT showed that Q0019+0107-type line profiles can be produced in a single,
rapidly rotating equatorial wind seen nearly edge-on (see lower left panel of Fig. \ref{fite2}). In the
same vein, in the bottom right panel of Fig. \ref{fite2}, we illustrate how the \ion{C}{iv} line profile of
Q0041-4023 can be reproduced by a two-component wind seen nearly pole-on. This arises when a single deep absorption
trough is associated with a quasi-symmetric emission profile. Once again, given the uncertainties
on the wind parameters because of the lack of clear signatures in the line shape, several wind models can be fitted to the observed line profiles (here we chose as typical values for model M2
$\rm{v}_{max}^{eq}=V_{rot}=0.5~\rm{v}_{max}^{po}$).

From Figs. \ref{fite1} and \ref{fite2}, we note that our simple model is able to reproduce the diversity
of the BAL profiles observed in a real sample of objects quite well. Interestingly enough, in order to be able to fit the \ion{C}{iv} profiles
with MCRT, we must shift the whole simulated line profile with respect to the emission peak, usually
used for redshift determination. This shift is needed to center the underlying emission component of the
profile on the zero velocity. Indeed, when dealing with resonant scattering, an absorption trough extending over the velocity range
[$-\rm{v}_{max}$, 0] produces an emission feature extending from $-\rm{v}_{max}$ to $+\rm{v}_{max}$. The
redshift of the quasar determined from the center of the underlying  \ion{C}{iv} emission line is given in the second
column of Table~\ref{tabfit}.

From the fitting procedure, one of the major parameters that control the line profile in this type of wind remains the viewing angle $i$, which plays a crucial role in the shape of
the absorption part of the line by controlling the relative contribution of the
equatorial and polar components (when both are required). Thus when a line profile exhibits a sharp, deep absorption trough superimposed on a shallower high-velocity absorption component,
the quasar is probably viewed along a line of sight, such as the dense equatorial wind seen
nearly edge-on (e.g. Q1413+117). However, it should be kept in mind that an edge-on disk does not necessarily
produce a completely black absorption trough, even for the highest optical depths (cf. Table \ref{tabfit}),
since photons scattered to the observer may not be completely reabsorbed by the disk then
filling in the absorption trough.
This agrees with what is usually observed among the Korista et al. (\cite{ko93}) spectra and has
already been pointed out by Lee \& Blandford (\cite{le97}) or Arav et al. (\cite{ar07}, and references therein).

Finally in all modeled profiles, we need to allow for the creation of photons inside the wind (a fraction $f_e > 0$ relative to the continuum intensity).
This intrinsic emission produces stronger emission peaks than in the case of a purely
resonant scattering wind.

\section{Discussion}

We showed that a simple two-component equatorial+polar wind model
is able to reproduce a variety of BAL profiles, ranging from detached
absorption troughs to P Cygni-type profiles. The solutions of the fits
are not unique and several models with different geometries and/or physical
properties can equally reproduce the observed spectra. In accordance with
previous studies (e.g. Hamann et al. \cite{ha93}), this demonstrates that
a unique physical characterization of the outflow cannot be derived from line profile fitting.

While detailed information on the geometry of the outflows cannot be derived, we nevertheless reached some
interesting conclusions. First, in some objects, it is necessary to include both the equatorial and the
polar absorption regions. This is indeed the case for objects like Q1413+117, Q1333+2840, but also
Q1235+1453 or Q0842+3431, where the polar component allows reproduction of the shallow absorption trough
observed at higher velocities and where the lower velocity equatorial component is needed to reabsorb
the emission from the polar wind.
In other objects (the prototype in our sample being Q0019+0107),
the polar component is not necessary and the profiles
can be fitted with only a rapidly rotating equatorial wind seen nearly
edge-on. In this particular case, there is a strong degeneracy between
the model parameters, as also suggested by similar
profiles having been computed in the framework of other wind models (e.g.
Proga \cite{pr03}, Proga \& Kallman \cite{pr04}).

Interestingly, in many of the fitted spectra, the viewing angle to the
wind axis is found to be high, suggesting that BAL quasars are
essentially observed when the optically thick equatorial wind blocks
the direct view to the continuum source. This agrees with the
high inclination generally inferred to account for the
spectropolarimetric properties of BAL QSOs (Schmidt \& Hines \cite{sc99}, Ogle et al. \cite{og99}). In this case, the polar wind could play the role of the
extended scattering region at the origin of the polarization.  We also
showed that, when the profile is constituted of a single deep
absorption trough and a quasi-symmetric emission peak
(e.g. Q0041-4023), the line can be produced in a two-component wind
seen at low inclination (see the bottom right panel of Fig.~\ref{fite2}).  This
result underlines that some BAL troughs can be observed as
the result of a polar outflow in a two-component wind with a large
covering factor.  As mentioned in the introduction, a growing number
of radio observations of BAL QSOs provide evidence for such
two-component wind models, while recent hydrodynamical simulations
show that a stable two-component wind can result from the flow around
a black hole and its accretion disk (e.g. Proga \cite{pr07}).  This model
appears challenging for the unification by orientation scheme of BAL
and non-BAL quasars (e.g. Turnshek \cite{tu84a},
Hamann et al. \cite{ha93}) because
of there is an absorption component at every viewing angle
(cf. panel E of Fig.~\ref{param}).  However, that the polar component
can be omitted in some cases (e.g. Q0019+0107) supports the existence
of a class of objects in which only a fraction of the continuum source
is covered by the BAL material. All in all, the observations suggest
that BAL outflows can have a wide range of covering factors.

In our study we decided to use
a simple radially expanding wind model with an equatorial and a polar component. This type of wind can account for most characteristics of the resonance line profiles observed
in the spectra of BAL QSOs.
However, we were not able to fit the \ion{C}{iv} line shape of the P Cygni-type quasar prototype PHL5200. This indicates that the model used does not include all the ingredients needed.
Indeed, the model cannot reproduce the very sharp transition observed between the absorption and the emission components
in the \ion{C}{iv} profile of PHL5200 (Turnshek et al. \cite{tuetal88}). Such a sharp transition at zero velocity
could in turn be produced in a wind launched from the disk itself. In that type of model, which exhibits large-scale properties similar to those of the model considered in the present study, the wind is launched from the disk and then radially accelerated by the radiation pressure (Murray et al. \cite{mu95}, Proga \& Kallman \cite{pr04}). When observed nearly edge-on, it can produce strong absorption at the center of the line as observed in PHL5200. Proga (\cite{pro03}) shows the capabilities of these winds to produce a sharp transition between the absorption and emission in the case of cataclysmic variable stars. Implementing that type of wind is beyond the scope of this paper and has been left for future work.

Furthermore, in a majority of objects we found it necessary to use high values of the ratio
of the rotational speed to the polar terminal speed of the wind $\rm{V}_{rot}/v_{max}^{po}$ to adequately reproduce the asymmetry of the emission line profiles. Such a high value of the rotation accounts for the redshift and the faintness of the emission peak. These characteristics result from the continuum emitted on the red side of the line profile possibly being absorbed by the optically thick material located between the source of continuum and the distant observer when rotation is present (see Fig.~\ref{figredabs}).
The rotational velocity at the base of the wind can reach several thousand $\rm{km~s}^{-1}$, which is consistent
with the rotational velocities inferred for gas orbiting in the vicinity of a supermassive black hole (e.g. Murray et al. \cite{mu95}, Young et al. \cite{yo07}). Interestingly, we found that the rotational
velocity must remain low in quasars with P Cygni-like line profiles. Indeed, to keep the emission
intense, the redward absorption due to the rotation must be small (cf.
Fig.~\ref{figredabs}). As a consequence, the shallow blue absorption wing only comes
from the polar wind. This may indicate a possible dynamical difference
between the BAL QSOs with detached profiles and those ones with P Cygni
type profiles.

The rotation of the wind naturally provides a simple and straightforward interpretation of the correlation
between the properties of the broad emission lines (BELs) and those of the broad absorption lines as reported by Turnshek (\cite{tu84a}). Indeed, as shown in that paper, BAL QSOs having complex absorption generally have a weaker \ion{C}{iv} emission peak
relative to the continuum than does BAL QSOs with smooth P Cygni absorption. This can be explained by the stronger/wider absorption on the red side of the line profile resulting from the rotation of optically thick material in front of
the continuum source.
Combining the emission profile and the redshifted absorption of the continuum results in profiles resembling the so-called detached troughs observed in several BAL QSO spectra (Korista et al. \cite{ko93}, Arav \& Begelman \cite{ar94}, Hall et al. \cite{ha02}, Proga \& Kallman \cite{pr04}). Recently, spectropolarimetry of the BAL QSO PG1700+518 have revealed significant variations in the polarization position angle over the H$\alpha$ BEL, which is interpreted as the typical signature of a rotating wind (Young et al. \cite{yo07}, Wang et al. \cite{wa07}).

\begin{table}
\caption{Blueshift of the \ion{C}{iv} line relative to the systemic redshift $z_{sys}$ for the three quasars from our sample with secured [\ion{O}{iii}] line measurement.}
\label{sysred}
\centering
\begin{tabular}{l l l l}     % 4 columns
\hline\hline
BAL QSO Name & $z_{sys}$ & Ref. $z_{sys}$ & $\Delta \rm{v_{blueshift}}$\\
\hline
 Q0019+0107 & 2.131 & Dietrich et al. \cite{di09} & 5500 $\rm{km}~s^{-1}$\\
 Q0226-1024 & 2.268 & McIntosh et al. \cite{mc99} & 9200 $\rm{km}~s^{-1}$\\
 Q1413+1143 & 2.553 & Hutsem\'ekers et al. \cite{hu09} & 2500 $\rm{km}~s^{-1}$\\
\hline
\end{tabular}
\end{table}

When fitting the \ion{C}{iv} line profiles, we found it necessary to adopt a systemic redshift
for the wind lower than the redshift given by the peak of the \ion{C}{iv} emission line.
This ``wind redshift" corresponds to the centro{\"\i}d of the full emission component
that underlies the observed absorption+emission profile and defines the rest frame
of the outflow model. It is interesting to compare this redshift to the redshift
determined from the narrow forbidden lines usually thought to provide the true
systemic redshift of the quasar and its host galaxy (e.g. McIntosh et al. \cite{mc99},
Vanden Berk et al. \cite{va01}). Unfortunately, only a few measurements are available
because the [\ion{O}{iii}] lines are shifted in the near-infrared and are
fainter in BAL QSOs than in non-BAL QSOs (Yuan et Wills \cite{yu03}). For the BAL QSOs
of our sample, only three accurate determinations are available, as given in Table~\ref{sysred}.
For these quasars, we find a net blueshift of the simulated \ion{C}{iv} line profiles by several thousand
$\rm{km}~s^{-1}$ with respect to the systemic redshift measured from [\ion{O}{iii}]. Such a blueshift of
the highly ionized species relative to the narrow [\ion{O}{iii}] lines --or to the low ionization \ion{Mg}{ii} line--
is rather common and well known in the case of both the BAL and non-BAL QSOs (e.g. Gaskell \cite{ga82},
Corbin \cite{co90}, McIntosh et al. \cite{mc99}), reaching up to 4000 $\rm{km}~s^{-1}$ (Corbin \cite{co90}).
BAL QSOs are among the QSOs with the highest blueshifts (Richards et al. \cite{ri02,ri08}), in agreement
with our measurements.
While the blueshift of the \ion{C}{iv} emission with respect to [\ion{O}{iii}] emission is well
documented, its origin is still unclear. Several mechanisms have been proposed to interpret it,
including dust attenuation of the red emission component, scattering of the line profile,
relativistic effects, and black hole recoil (cf. Corbin \cite{co90}, Mc Intosh et al. \cite{mc99}, Vanden Berk
et al. \cite{va01}, Shields et al. \cite{sc09}). Ultimately, it might be necessary to consider these effects for
full self-consistent modeling of quasar outflows.
In the particular case of Q0019+0107, two narrow \ion{C}{iv} absorption lines are observed at -3000 and
+3000 $\rm{km}~s^{-1}$ in the wind rest frame (Fig. 4), i.e. at -8000 and -2000 $\rm{km}~s^{-1}$
in the systemic quasar+host rest frame. These velocities suggest that they originate in a large-scale
outflow in the host galaxy rather than in the BAL wind. Interestingly enough, the velocity
difference between these narrow absorption lines is close to the velocity separation between the
Ly$\alpha$ and NV resonance doublets ($\sim 5900$ $\rm{km}~s^{-1}$), suggesting a possible line
locking effect (Korista et al. \cite{ko93}, Arav \& Begelman \cite{ar94}).

\section{Conclusions}

In this study, we used a combination of a Monte Carlo radiative transfer code and a simple two-component polar+equatorial wind model in which photons are emitted from a central spherically symmetric source
and resonantly scattered in the wind to reproduce typical \ion{C}{iv} resonance line profiles
selected from a homogeneous sample of BAL QSO spectra.

Although the lack of uniqueness of the line profile fitting does not allow us to strongly
constrain the geometry of the wind, we can summarize our main findings as follows
\begin{enumerate}
      \item The diversity of BAL profiles produced by the adopted polar+equatorial model ranges from the
            typical P Cygni-type profiles to the detached absorption ones, reproducing those observed in a homogeneous
            sample of BAL QSOs.
      \item While in some cases the line profiles can be reproduced by a single equatorial wind,
            we find it necessary to use a two-component polar+equatorial wind in a majority of objects.
      \item The viewing angle to the wind is generally large (disk seen near edge-on); however
            in some cases, the line profiles can also be reproduced when assuming a pole-on view, in
            accordance with the results of recent radio surveys of BAL QSOs. In this context, it
            would be interesting to obtain good quality spectra of bona-fide polar BAL quasars
            and try to fit their line profiles by assuming the pole-on view.
      \item The equatorial wind is rotating, and the rotational velocity at the base of the wind can reach a significant fraction of the polar terminal speed.
   \end{enumerate}

A possible way to break the degeneracy between the various parameter combinations of the
two-component model that can reproduce the observed BAL profiles
is to use gravitational microlensing. Indeed, a microlens moving
across the quasar inner regions can differentially
magnify the different line-forming regions, inducing line profile variations from which the
geometry of the outflow can in principle be retrieved (e.g. Hutsem\'ekers \cite{hu93}, Hutsem\'ekers et al. \cite{hu94}, Lewis \& Belle \cite{le98}, Chelouche \cite{ch03,ch05}). Our code MCRT has been explicitly built
to integrate these microlensing effects. The effect of microlensing on BAL profiles, their
use for deriving the physical properties of the outflow, and application to a known
lensed system will be presented in a second paper.

%
%______________________________________________________________

%\begin{acknowledgements}
%      Part of this work was financially supported a Ph.D. student grant of the Belgian Fund for Scientific Research (F.N.R.S.).
%This work was also supported in part by the PRODEX Experiment Agreement 90195 (ESA and PPS Science Policy, Belgium).
%\end{acknowledgements}

\end{document}